\documentstyle[12pt]{article}
\def\ie{{\em i.e.}}
\def\eg{{\em e.g.}}

\def\beq{\begin{equation}}
\def\eeq{\end{equation}}

\catcode`\@=11 % This allows us to modify PLAIN macros.

\def\lsim{\mathrel{\mathpalette\@versim<}}
\def\gsim{\mathrel{\mathpalette\@versim>}}
\def\@versim#1#2{\vcenter{\offinterlineskip
    \ialign{$\m@th#1\hfil##\hfil$\crcr#2\crcr\sim\crcr } }}

\def\JL{J. L. Lopez}
\def\DVN{D. V. Nanopoulos}

\def\r#1{$\bf#1$}
\def\rb#1{$\bf\overline{#1}$}

\def\t1{{\tilde 1}}

\def\GeV{\,{\rm GeV}}

\def\to{\rightarrow}

\def\NPB#1#2#3{Nucl. Phys. B {\bf#1} (19#2) #3}
\def\PLB#1#2#3{Phys. Lett. B {\bf#1} (19#2) #3}

\def\PRD#1#2#3{Phys. Rev. D {\bf#1} (19#2) #3}

\def\PRT#1#2#3{Phys. Rep. {\bf#1} (19#2) #3}

\def\TAMU#1{Texas A \& M University preprint CTP-TAMU-#1}

\begin{document}

% TH format
\begin{flushright}
\baselineskip=12pt
{CTP-TAMU-06/94}\\
{ACT-02/94}\\
\end{flushright}

\begin{center}
\vglue 0.6cm
{\Large\bf Closing the Universe by relaxing\\
 the cosmological constant\\}
\vspace{0.2cm}
\vglue 1cm
{JORGE L. LOPEZ and D. V. NANOPOULOS\\}
\vglue 0.4cm
{\em Center for Theoretical Physics, Department of Physics, Texas A\&M
University\\}
{\em College Station, TX 77843--4242, USA\\}
and\\
{\em Astroparticle Physics Group, Houston Advanced Research Center
(HARC)\\}
{\em The Mitchell Campus, The Woodlands, TX 77381, USA\\}
\baselineskip=12pt

\vglue 1cm
{\tenrm ABSTRACT}
\end{center}
%\vglue -0.2cm
{\rightskip=3pc
 \leftskip=3pc
%\xpt\baselineskip=12pt
\noindent
We propose a string-inspired model which correlates several aspects of
particle physics and cosmology. Inspired by the flat directions and the absence
of adjoint Higgs representations found in typical string models, we consider a
no-scale $SU(5)\times U(1)$ supergravity model. This model entails well
determined low-energy phenomenology, such as the value of the neutralino dark
matter relic abundance and a negative contribution to the vacuum energy. A
positive contribution to the vacuum energy is also typically present in string
theory as a consequence of the running of the fundamental constants towards
their fixed point values. If these two contributions  cancel appropriately, one
may end up with a vacuum energy which brings many cosmological observations
into better agreement with theoretical expectations. The present abundance of
neutralinos would then be fixed.  We delineate the regions of parameter space
allowed in this scenario, and study the ensuing predictions for the sparticle
and Higgs-boson masses in this model.
}
% TH format
\vspace{1cm}
\begin{flushleft}
\baselineskip=12pt
{CTP-TAMU-06/94}\\
{ACT-02/94}\\
January 1994
\end{flushleft}
\vfill\eject
\setcounter{page}{1}
\pagestyle{plain}

\baselineskip=14pt

There are two fundamental problems in cosmology today which are intimately
related to particle physics: the dark matter problem and the cosmological
constant problem. Indeed, almost any extension of the minimal Standard Model
($SU(3)\times SU(2)\times U(1)$) predicts some new particles which can be
identified as viable candidates for hot (\eg, massive neutrinos and axions) or
cold (\eg, neutralinos and cryptons) dark matter. Concerning the cosmological
constant problem, there are not that many satisfactory solutions. Recent
developments at the cosmological and particle physics fronts indicate a
possible correlation between dark matter and the cosmological constant
($\Lambda$). In this note we allow the cosmological constant to contribute to
the cosmic energy density, thus reducing the corresponding dark matter
contribution such that the latest cosmological observations are best fit.

The standard cold dark matter model (with $\Omega=1$ and $h=0.5$) has had great
success, but as observations have improved various discrepancies with the
data have started to appear \cite{Davis}. Recent observations of X-rays from
the gas surrounding galaxy clusters indicate that $\Omega$ (visible plus
dark) is ${\cal O}(0.2)$, implying $\Omega_{\rm CDM}\lsim0.2$ \cite{BS}.
On the other hand, an $\Omega=1$ Universe is not only appealing in the
inflationary scenario, but it may also arise under more general circumstances
(as we discuss below). Thus, the phenomenological suggestion has been made that
a Universe with $\Omega_{\rm CDM}\approx0.2$, $\Omega_{\Lambda}\approx0.8$,
and $h\approx1$ should be seriously considered \cite{ESM}. Indeed, such a
cosmological scenario is claimed to be consistent with a varied set of
cosmological observations
\cite{CE}: the APM galaxy angular correlation function, the rich cluster
correlation function, the mass function and power spectra from clusters, the
CfA slices, and the Southern Sky redshift survey.

The presently unique candidate for the unification of all fundamental
interactions, including gravity, is string theory. There are a few generic
properties of string theory which are relevant to our cosmological scenario.
First, string theory is characterized by flat directions, which implies that
certain parameters of the theory can only be determined dynamically at the
quantum level. This fundamental property of string theory \cite{Witten}
appears from the low-energy effective field theory point of view as the
so-called {\em no-scale} structure \cite{LN}. This mechanism can explain the
smallness of the electroweak scale (\ie, $M_W={\cal
O}(e^{-1/\alpha_t})M_{Pl}$), and yields a negative cosmological constant ${\cal
O}(-M_W^4)$ (much smaller than the usual $M_W^2M^2_{Pl}$ or $M^4_{Pl}$
results). Second, the simplest and most studied string models (\ie, those based
on level-one Kac-Moody algebras) do not contain adjoint Higgs representations
at the field theory level \cite{ELN}. This result precludes the traditional
grand unified gauge groups (\ie, $SU(5),SO(10),E_6$), but allows the
$SU(5)\times U(1)$ or flipped $SU(5)$ gauge group, since it only requires the
allowed \r{10},\rb{10} representations for symmetry breaking down to the
Standard Model gauge group \cite{revitalized}. There are many interesting
properties of $SU(5)\times U(1)$ which do not impact directly on our present
discussion, but when combined with the no-scale structure result in rather
restricted three-parameter models (compared with the more than 20 parameters of
the Minimal Supersymmetric Standard Model) \cite{EriceDec92}. As a result, the
dependence of the neutralino relic abundance $\Omega_\chi$ on the few model
parameters can be studied in detail. It is important to point out
that a cosmologically desirable value of $\Omega_\chi h^2$ (\ie, $\Omega_\chi
h^2\lsim1$) is not a guaranteed prediction of supersymmetric models, as values
of $\Omega_\chi h^2$ as large as 100 are not uncommon \cite{LNYdmI+KLNPYdm}.
However, in the three-parameter models which we consider, $\Omega_\chi h^2<1$
is automatic.

Finally, it has been recently argued that string theory in an expanding
Universe cannot be considered to be in ``equilibrium", but that
it rather ``runs" towards a fixed point \cite{EMN}. The renormalization group
scale which characterizes this ``running" is identified as a
statistically-defined universal ``time". The arrow of time is thus explained.
This scenario entails that all {\em dimensional} fundamental constants of
Nature actually ``run" until they reach the equilibrium (fixed) point of this
flow. The cosmological constant obeys a simple relaxation equation
$\Lambda(t)=\Lambda(0)/[1+t\Lambda(0)]$, where all quantities are given
in Planck (``natural") units (\ie, $t\to t M_{Pl}$, $\Lambda\to\Lambda/
M^2_{Pl}$, $G_N=1$). The vacuum energy is then given by $\rho_{\rm
VAC}=\Lambda/8\pi$.

 For the present epoch $t_0\sim10^{60}$, and with $\Lambda(0)\sim 1$ one gets
$\Lambda(t_0)\sim10^{-60}$. Combining our above discussion on the no-scale
supergravity prediction for the standard contribution to the vacuum energy
(\ie, $\rho_{\rm VAC}\sim -(M_W/M_{Pl})^4\sim-10^{-60}$) with the
string-theoretic contribution (\ie, $\rho_{\rm VAC}\sim10^{-60}$), it may be
possible to obtain a cancellation of the total vacuum energy. This argument is
only suggestive, although encouraging. Once the effective field theory from
string is completely understood, it should be possible to explain what symmetry
principle is behind the phenomenological requirement of $\rho^{\rm tot}_{\rm
VAC}\approx0$. Note that since the Universe has yet to relax completely to its
true vacuum state, it is possible that $\rho^{\rm tot}_{\rm VAC}(t_0)\sim{\cal
O}(10^{-123})$, in agreement with present experimental limits,\footnote{For the
$\Omega_\Lambda=0.8$, $h=1$ cosmological constant model, $\rho_{\rm
VAC}=3\times 10^{-123}$.} and yet eventually it would relax to $\rho^{\rm
tot}_{\rm VAC}(t_{\rm equil})\approx0$.

Interestingly enough, this dynamical out-of-equilibrium $\rightarrow$
equilibrium string scenario provides an appealing alternative to conventional
inflation, resolving the standard cosmological problems (horizon, flatness,
large entropy) automatically \cite{EMN}. For instance, the running of the
fundamental constants implies that the speed of light $c$ becomes infinite in
the early Universe, thus solving the horizon problem. For our present purposes,
$\Omega=1$ is then a string-derived property.

In view of the above discussion, we then propose a string-inspired cosmological
model: no-scale $SU(5)\times U(1)$ supergravity with $\Omega_\Lambda=0.8$ and
$h=1$, as preferred by cosmology, and $\Omega_{\rm CDM}=0.2$ so that
$\Omega=\Omega_\Lambda+\Omega_{\rm CDM}=1$, as preferred by string theory.

For practical purposes, the most important feature of no-scale $SU(5)\times
U(1)$ supergravity is the minimality of parameters needed to describe the
complete low-energy supersymmetric spectrum and its interactions. The
constraints of supergravity and radiative electroweak symmetry breaking reduce
the number of parameters to four: the ratio of Higgs-boson vacuum expectation
values ($\tan\beta$) and three universal soft-supersymmetry breaking parameters
($m_{1/2},m_0,A$). In addition, the top-quark mass ($m_t$) plays an important
role through the running of the mass parameters from the unification scale down
to the electroweak scale. In no-scale $SU(5)\times U(1)$ supergravity
\cite{EriceDec92} we consider two string-inspired scenarios for the {\em
universal} soft-supersymmetry-breaking parameters: (i) the {\em moduli}
scenario \cite{Lahanas+EKNI+II}, where $m_0=A=0$, and (ii) the dilaton scenario
\cite{KL+Ibanez}, where $m_0={1\over\sqrt{3}}m_{1/2}$, $A=-m_{1/2}$. In this
case, the number of parameters is just two ($\tan\beta,m_{1/2}$) plus the
top-quark mass. For the typical value of $m_t=150\GeV$ we find the following
allowed range of $\tan\beta$: $2\lsim\tan\beta\lsim26\,(40)$ in the moduli
(dilaton) scenario. The resulting parameter space for the moduli \cite{LNZI}
and dilaton \cite{LNZII} scenarios consists of discrete pairs of points in the
$(\tan\beta,m_{1/2})$ plane. In practice, we trade the $m_{1/2}$ supersymmetric
mass scale parameter for the more readily measurable lightest chargino mass
$m_{\chi^\pm_1}$. An extensive study of the various experimental constraints
and experimental predictions in $SU(5)\times U(1)$ supergravity has been
recently given in Ref.~\cite{Easpects}. We refer the reader to that reference
for further details.

The computation of the neutralino relic abundance, following the methods of
Ref.~\cite{LNYdmI+KLNPYdm}, shows that $\Omega_\chi h^2\lsim0.25\,(0.90)$ in
the moduli (dilaton) scenario. On the other hand, the cosmological model with a
non-zero cosmological constant fits observations best for $\Omega_{\rm
CDM}\approx0.2$ and $h\approx1$, that is $\Omega_\chi h^2\approx0.2$.
This constraint can be applied to the still-allowed parameter space in both
moduli and dilaton scenarios. In Fig.~\ref{Figure1} we show the still-allowed
points in parameter space \cite{Easpects}. The areas of the plots which do not
contain any points are theoretically or experimentally excluded. In the figure
we have delineated the region where $\Omega_\chi h^2=0.2$ (the boundary between
the pluses $+$ and the diamonds $\diamond$). The cosmological constraint then
allows one to determine $\tan\beta$ for a given chargino mass.
Moreover, we can compute the sparticle and Higgs-boson masses along this
boundary, \ie, when the relation $\Omega_\chi h^2=0.2$ is satisfied. These
are shown in Fig.~\ref{Figure2} for the moduli and dilaton scenarios. The
resulting spectra indicate that the gluino, squarks, sleptons, and charginos
should be accessible at the LHC, but not at LEPII. On the other
hand, the lightest Higgs boson ($h$) may be accessible at LEPII
\cite{Easpects}.

In sum, our proposed string-inspired model correlates several aspects of
particle physics and cosmology. Inspired by the string flat directions and the
absence of adjoint Higgs representations, we have considered a no-scale
$SU(5)\times U(1)$ supergravity model. This entails well determined low-energy
phenomenology, such as the value of the neutralino dark matter relic abundance
and a negative contribution to the vacuum energy. A positive contribution to
the vacuum energy is also typically present in string theory as a consequence
of the running of the fundamental constants towards their fixed point values.
If these two contributions cancel appropriately, one may end up with a vacuum
energy which brings many cosmological observations into better agreement with
theoretical expectations. The present abundance of neutralinos would therefore
be fixed and this entails clear predictions for the sparticle and Higgs-boson
masses in the model.

\section*{Acknowledgements}
This work has been supported in part by DOE grant DE-FG05-91-ER-40633.

%\newpage

\section*{Figure Captions}
\begin{enumerate}
\item
The still-allowed points in the parameter space of no-scale $SU(5)\times U(1)$
supergravity in the (a) moduli and (b) dilaton scenarios, for $m_t=150\GeV$.
The cosmological constraint $\Omega_\chi h^2=0.2$ is satisfied along the
boundary between the pluses ($+$) and the diamonds ($\diamond$).
\label{Figure1}
\item
The predicted values for the gluino, squark, slepton, and ligthest Higgs-boson
masses in no-scale $SU(5)\times U(1)$ supergravity in the moduli and dilaton
scenarios, when the cosmological constraint $\Omega_\chi h^2=0.2$ is
satisfied.
\label{Figure2}
\end{enumerate}

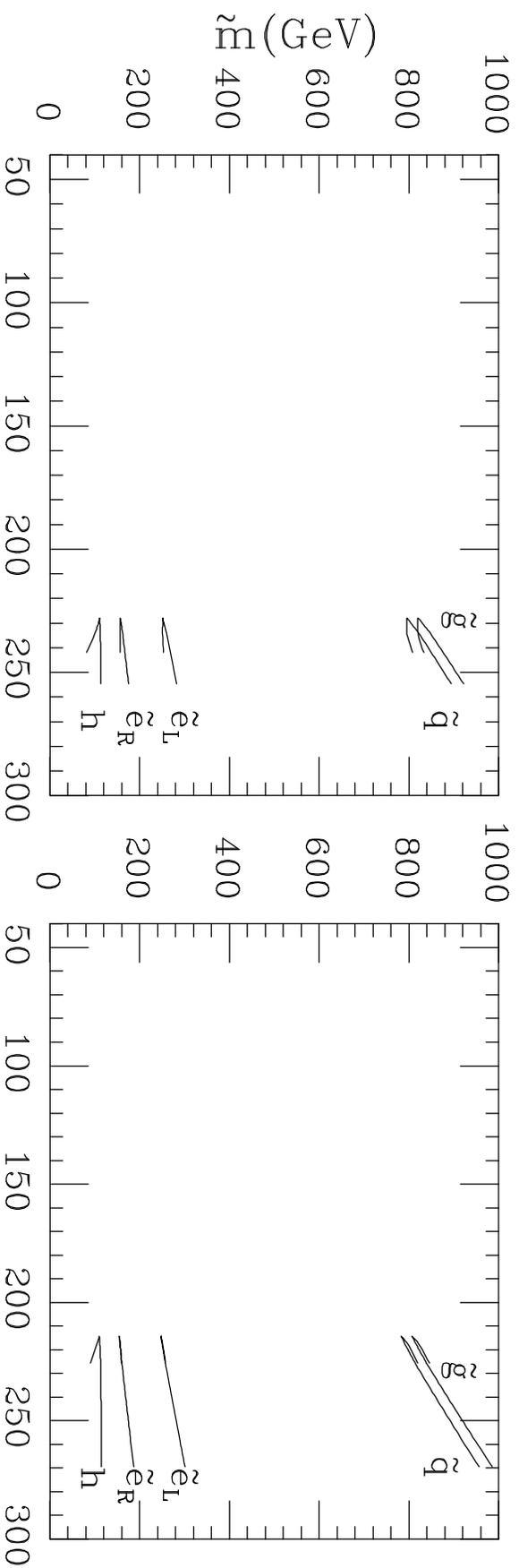
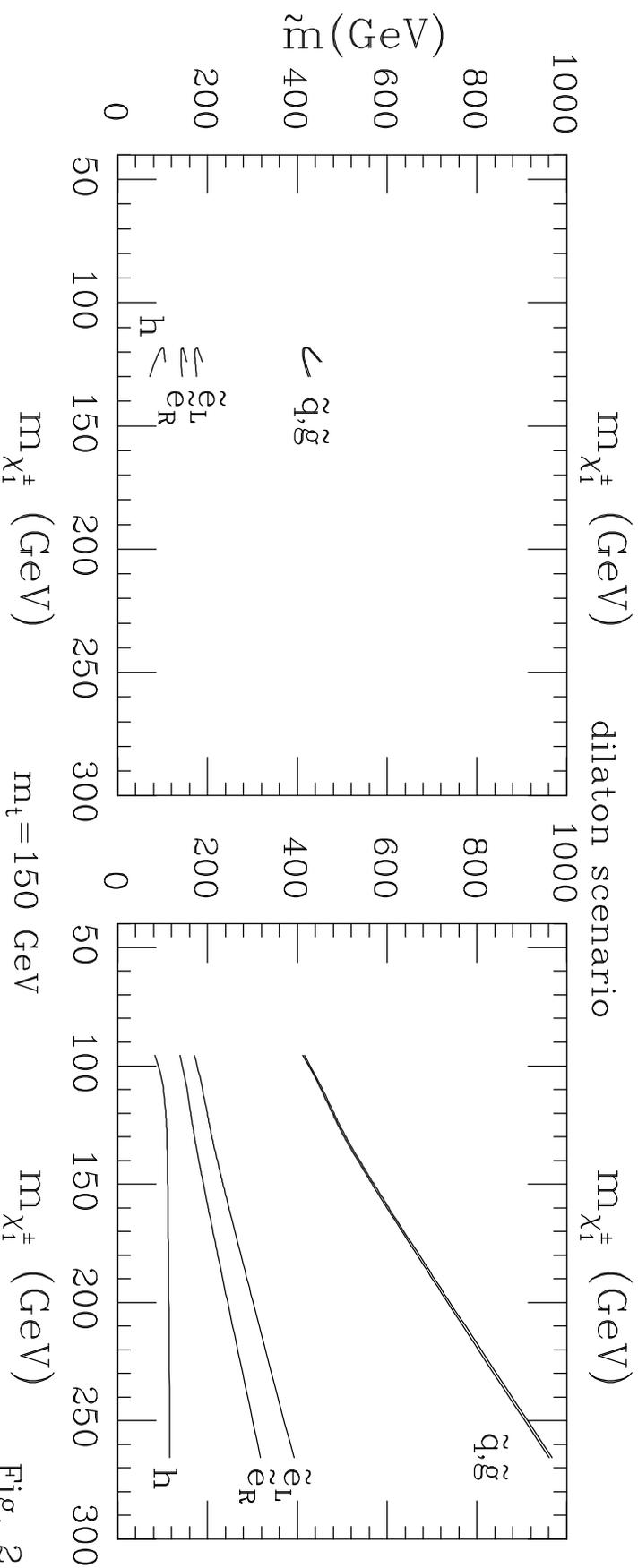
\end{document}